\documentclass[12pt]{article}
\usepackage{epsfig}
\begin{document}
\title{The Domino Effect}
\author{J. M. J. van Leeuwen\\
Instituut--Lorentz, Leiden University,  P. O. Box 9506, \\
2300 RA  Leiden, The Netherlands} 
\date{\today}
\maketitle

\begin{abstract}
The physics of a row of toppling dominoes is discussed. In particular the forces between
the falling dominoes are analyzed and with this knowledge, the effect of friction has been
incorporated. A set of limiting situations is discussed in detail, such as the limit of
thin dominoes, which allows a full and explicit analytical solution. The propagation speed
of the domino effect is calculated for various spatial separations. Also a
formula is given, which gives explicitly the main dependence of the speed as function of the 
domino width, height and interspacing.
\end{abstract}

\section{Introduction}

Patterns formed by toppling dominoes are not only a spectacular view, but their
dynamics is also a nice illustration of the mechanics of solid bodies.
One can study the problem on different levels. Walker \cite{walker} gives a qualitative 
discussion. Banks \cite{banks}
considers the row of toppling dominoes as a sequence of 
independent events: one domino undergoes a free fall, till it hits the next one, which
then falls independently of the others, and so on. He assumes that in the collision the 
linear momentum along the supporting table is transmitted. 
This is a naive viewpoint, but it has the advantage
that the calculation can be carried out analytically. A much more thorough treatment
has been given by D. E. Shaw \cite{shaw}.
His aim is to show that the problem is a nice illustration of computer aided instruction 
in mechanics. He introduces the basic feature that the domino, after having struck the next one,
keeps pushing on it. So the collision is completely {\em inelastic}. 
In this way a train develops of dominoes leaning on each other and
pushing the head of the train. One may see this as an elementary demonstration of a
propagating soliton, separating the fallen dominoes from the still upright ones.
Indeed Shaw's treatment is a simple example how to handle holonomous constraints 
in a computer program describing the soliton. As collision law he takes conservation
of angular momentum. We will demonstrate, by analyzing the forces between the dominoes,
that this is not accurate. The correction has a substantial influence on the solition speed,
even more important than the inclusion of friction, which becomes possible when the
forces between the dominoes are known. 

The setting is a long row of identical and perfect dominoes of height $h$, 
thickness $d$ and interspacing $s$. In order to make the problem tractable we assume
that the dominoes only rotate (and e.g.~do not slip on the supporting table). Their fall
is due to the gravitational force, with acceleration $g$. The combination $\sqrt{gh}$
provides a velocity scale and it comes as a multiplicative factor in the soliton speed.
Typical parameters of the problem are the aspect ratio $d/h$, which is determined by the 
type of dominoes used, and the ratio $s/h$, which can be easily varied in an experiment. 
Another characteristic of the dominoes is their mutual  friction coefficient $\mu$ which
is a small number ($\sim 0.2$). The first domino gets a gentle push, such that it topples and
makes a ``free rotation'' till it strikes the second. After the collision the two fall together till 
they struck the third and so forth. So we get a succession of rotations 
and collisions, the two processes being governed by different dynamical laws. Without
friction the rotation conserves energy, while the constraints exclude the energy to be
conserved in the collision. In fact this is the main dissipative element, more than the 
inclusion of friction.

The goal is to find the dependence of the soliton speed on the interdistance $s/h$.
In the beginning this speed depends on the initial push, but
after a while a stationary pattern develops: a propagating soliton with upright dominoes
in front and toppled dominoes behind. The determination of the forces between the
dominoes requires that we first briefly outline the analysis of Shaw. Then we analyze the
forces between the dominoes. Knowing these we make the collision law more
precise. With the proper rotation and collision laws we give the equations for the fully 
developed solitons. 
The next point is the introduction of friction and the calculation of its effect
on the soliton speed. As illustration we discuss the limit of thin dominoes $d \rightarrow 0$,
with permits for small interseparations a complete analytical solution. Finally we present
our results for the asymptotic soliton speed for various values of the friction and compare 
them with some experiments.  We also give an explicit formula, which displays the main 
dependence of the soliton speed on the parameters of the problem. The
paper closes with a discussion of the results and the assumptions that we have made.

\section{Constraints on the Motion} \label{constraints}

\begin{figure}[h]
\begin{center}
    \epsfxsize=12cm
    \epsffile{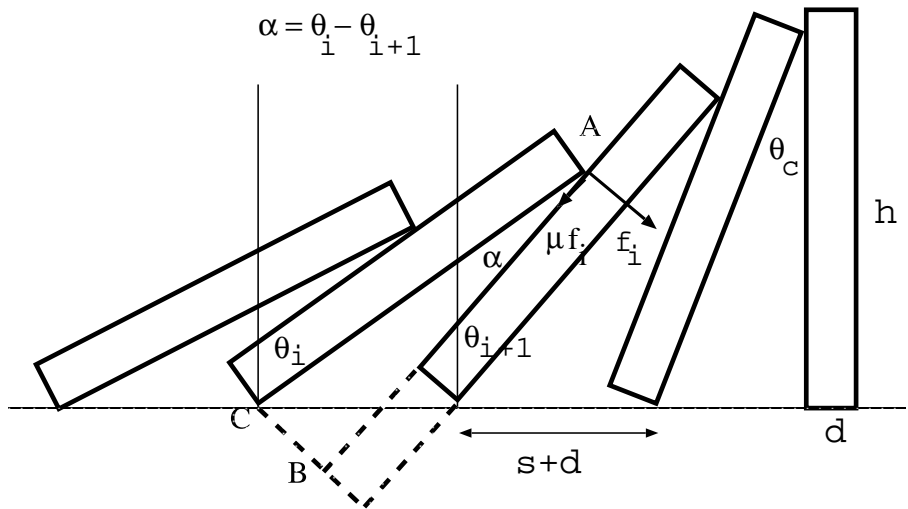}
    \caption{Successive dominoes. The tilt angle $\theta_i$ is taken with respect to 
the vertical. In the rectangular traingle ABC the top angle is $\alpha = \theta_i - \theta_{i+1}$,
the hypotenuse has the length $h$ and the base BC the length $(s+d) \cos \theta_{i+1} - d$.
Expressing this base in the hypotenuse and the top angle yields relation (\ref{a1}). In the
picture the tilt angle of the head of chain $\theta_n$ has reached its final 
value $\theta_c = \arcsin(s/h)$. The first domino has almost reached the stacking angle 
$\theta_\infty$. The normal force $f_i$ and the friction force $\mu f_i$ that domino $i$ 
exerts on $i+1$ are also indicated.}  \label{dominoes}
\end{center}
\end{figure}

The basic observation is that domino $i$ pushes over domino $i+1$ and 
{\it remains in contact afterwards}. So after the contact of $i$ with $i+1$ the motion of
$i$ is constrained by the motion of $i+1$. Therefore we can take the tilt angle $\theta_n$ of 
the foremost falling domino, as the only independent mechanical variable  
(see Fig. \ref{dominoes}). Simple goniometry tells that
\begin{equation} \label{a1}
h \sin(\theta_i - \theta_{i+1}) = (s+d) \cos \theta_{i+1} - d.
\end{equation} 
To see this relation it helps to displace domino $i+1$ parallel to itself, till its bottom line
points at the rotation axis of domino $i$ (see Fig. \ref{dominoes}). By this relation one can 
express the tilt angle $\theta_i$ in terms of the next $\theta_{i+1}$ and so on, such that all 
preceding tilt angles are expressed in terms of $\theta_n$.
The recursion defines $\theta_i$ as a function of $\theta_n$ of the form
\begin{equation} \label{a2}
\theta_i = p_{n-i} (\theta_n),
\end{equation}
i.e. the functional dependence on the angle of the head of the train depends only on the
distance $n-i$. The functions $p_j (\theta)$ satisfy
\begin{equation} \label{a3}
p_j (\theta) = p_{j-1} (\theta)) + \arcsin \left( {(s+d) \cos p_{j-1} (\theta) - d \over h} 
\right),
\end{equation} 
with the starting function $p_0 (\theta) = \theta$. They are defined on the interval
$0< \theta < \theta_c$, where $\theta_c$ is the angle of rotation at which the head of the
train hits the next domino
\begin{equation} \label{a4}
\theta_c = \arcsin (s/h).
\end{equation} 
We will call $\theta_c$ the angular distance. From the picture it is clear that the functions
are bounded by the value $\theta_\infty$, which is the angle for which the right hand side of
(\ref{a1}) vanishes
\begin{equation} \label{a5}
\cos \theta_\infty = {d \over s+d}.
\end{equation} 
$\theta_\infty$ is the angle at which the dominoes are stacked against each other at the
end of the train. We call $\theta_\infty$ the stacking angle.

The picture shows that the functions $p_j (\theta)$ are monotonically increasing functions.
They become flatter and flatter with the index $j$ and converge to the value $\theta_\infty$
(at least not too close to the maximum separation $s=h$, see Section \ref{calc}).
The functions are strongly interrelated, not only by the defining recursion
(\ref{a3}). The angle $\theta_i$ can be calculated from the head of the train $\theta_n$
by $p_{n-i}$ but also from an arbitrary intermediate $\theta_k$ by $p_{k-i}$. This implies
\begin{equation} \label{a6}
p_{n-i} (\theta) = p_{k-i} (p_{n-k} (\theta)), \quad \quad \quad {\rm e.g.}
\quad \quad \quad p_j (\theta) = p_{j-1} (p_1 (\theta)).
\end{equation} 
One easily sees that $p_1 (0) = \theta_c$. Therefore one has
\begin{equation} \label{a7}
p_j (0) = p_{j-1} (p_1 (0) ) = p_{j-1} ( \theta_c),
\end{equation} 
a property that will be used later on several times.

An inmediate consequence of (\ref{a1}) is the expression for the angular velocities
$\omega_i = d \theta / dt$ in terms of $\omega_n$. From the chain rule of differentiation 
we find
\begin{equation} \label{a8}
\omega_i = {d \theta_i \over d \theta_n} \, {d \theta_n \over dt} = w_{n-i} \, \omega_n,
\end{equation} 
with 
\begin{equation} \label{a9}
w_j (\theta) = {d p_j (\theta) \over d \theta}. 
\end{equation} 
Computationally it is easier to calculate the $w_j$ recursively. Differentiation of (\ref{a3})
with respect to $\theta_n$ yields
\begin{equation} \label{a10}
w_j (\theta) = w_{j-1} (\theta) \left(1 - {(s+d) \sin p_j (\theta)
\over h \cos\, [\,p_j (\theta) - p_{j-1} (\theta)]} \right).
\end{equation}    
Another useful relation follows from differentiation of the second relation (\ref{a6})
\begin{equation} \label{a11}
w_j (\theta) = w_{j-1} (p_1 (\theta)) \,w_1 (\theta) \quad \Rightarrow \quad 
w_j (0) = w_{j-1} (\theta_c),
\end{equation}
since $p_1 (0) = \theta_c$ and $w_1 (0) = 1$.
 
\section{Rotation Equations} \label{motion}

Without friction, the 
motion between two collisions is governed by conservation of energy, which consists
out of a potential and a kinetic part. The potential part derives from the combined height 
of the center of mass of the falling dominoes, for which we take the dimensionless quantity
\begin{equation} \label{b1}
H_n (\theta_n) = \sum^n_i [ \cos \theta_i + (d/h ) \sin \theta_i].
\end{equation} 
The kinetic part is given by the rotational energy, for which holds
\begin{equation} \label{b2}
K_n (\theta_n, \omega_n) = (I/2)  \sum^n_i \omega^2_i, \quad \quad \quad I = 
(1/3) m (h^2 + d^2), 
\end{equation} 
where $I$ is the angular moment of inertia with respect to the rotation axis and $m$ is the 
mass of the dominoes. We write the total energy as
\begin{equation} \label{b3}
E_n = {1 \over 2} m g h \, e_n = {1 \over 2} m g h \left( H_n (\theta_n ) + 
{I \over mgh} I_n (\theta_n)  \, \omega^2_n \right),
\end{equation} 
where the dimensionless effective moment of inertia $I(\theta_n)$ is defined as
\begin{equation} \label{b4}
I_n(\theta_n) = \sum^n_j \, w^2_j (\theta_n).
\end{equation} 
We have factored out $mgh/2$ in (\ref{b3}) as it is an irrelevant energy scale. This has the 
advantage that the expression between brackets is dimensionless. The factor $I/mgh$ 
\begin{equation} \label{b4a}
{I \over mgh} = {h (1 + d^2/h^2) \over 3g}
\end{equation} 
provides a time scale that can be incorporated in $\omega_n$. From now on we put this 
factor equal to unity in the formulae and remember its value when we convert 
dimensionless velocities to real velocities. 

We see (\ref{b3}) as the defining expression for $\omega_n$ as function of $\theta_n$
\begin{equation} \label{b5}
\omega_n (\theta_n) = \left({e_n - H_n(\theta_n )) \over I_n(\theta_n)}\right)^{1/2}.
\end{equation} 
As mentioned $e_n$ is a constant during interval $n$. So we can solve the temporal
behavior of $\theta_n$ from the equation
\begin{equation} \label{b6}
{d \theta_n (t) \over d t } = \omega_n (\theta_n).
\end{equation} 
The initial value for $\theta_n$ is 0 and the final value equals the rotational
distance $\theta_c $. The duration of the 
time interval where $n$ is the head of the chain, follows by integration
\begin{equation} \label{b7}
t_n = \int^{\theta_c}_0 {d \theta_n  \over \omega_n(\theta_n)}.
\end{equation} 
In this time interval the soliton has advanced a distance $s+d$. The ratio $(s+d)/t_n$ gives
the soliton speed, when the head of the train is at $n$. In order to integrate
the equations of motion (\ref{b6}) we must have a value for $e_n$ which basically amounts
to finding an initial value $\omega_n(0)$ as one sees from (\ref{b3}). In the next section
we outline how to calculate successively the $\omega_n (0)$.

Putting all ingredients together we obtain the asymptotic soliton speed $v_{\rm as}$ as
\begin{equation} \label{b8}
v_{\rm as} = \sqrt{gh} \left({3 \over 1 + d^2/h^2} \right)^{1/2} {s+d \over h} 
\lim_{n \rightarrow \infty} {1 \over t_n}.
\end{equation} 
In this formula the time $t_n$ is computed from the dimensionless equations (setting
$I/mgh$ equal to 1).

\section{The Collision Law, first version} \label{collision}

We now investigate what happens when domino $n$ hits $n+1$. In a very short time
domino $n+1$ accumulates an angular velocity $\omega_{n+1} (0)$. The change in 
$\omega_{n+1}$ takes place while the tilt angles of the falling dominoes  hardly change. 
Shaw \cite{shaw} postulates that the total angular momentum of the system is unchanged 
during the collision. This is not self-evident and we comment on it in Section \ref{coltrain}. 
Before the collision we have the angular momentum
\begin{equation} \label{c1}
L_n = \sum^n_j w_j (\theta_c) \, \omega_n (\theta_c). 
\end{equation} 
After the collision we have
\begin{equation} \label{c2}
L_{n+1} = \sum^{n+1}_j w_j (0) \, \omega_{n+1} (0).
\end{equation} 
Equating these two expressions yields the relation
\begin{equation} \label{c3}
\omega_{n+1} (0) = \omega_n (\theta_c) \, \sum^n_j w_j (\theta_c) / 
\sum^{n+1}_j w_j (0).
\end{equation} 
With the aid of this value we compute the total energy $e_{n+1}$ and the next integration
can be started. 
For the first time interval holds $e_0= 1 + \omega^2_0(0)$ since only the zeroth domino 
is involved and it starts in upright position
with angular velocity $\omega_0 (0)$. The value of $\omega_0 (0)$ has no influence on the
asymptotic behavior. After a sufficient number of time intervals, a stationary soliton develops.

\section{Forces between the Dominoes} \label{forces}

Conservation of energy requires the dominoes to slide frictionless over each other. 
Before we can introduce friction we have to
take a closer look at the forces between the falling dominoes. Without friction
the force which $i$ exerts on $i+1$ is perpendicular to the surface of $i+1$ with a 
magnitude $f_i$ (see Fig. \ref{dominoes}). Consider to begin with the head of the train $n$.
Domino $n$ feels the gravitational pull with a torque $T_n$ 
\begin{equation} \label{d1} 
T_n = (\sin \theta_n - (d/h) \cos \theta_n)/2,
\end{equation}
and a torque from domino $n-1$ equal to the force $f_{n-1}$ times the moment arm with 
respect to the rotation point of $n$. The equation of motion for $n$ becomes
\begin{equation} \label{d2}
{d \omega_n \over dt} = T_n + f_{n-1} \, h [\, \cos (\theta_{n-1}- \theta_n)- 
(s+d) \sin \theta_{i+1}\,].
\end{equation} 
Domino $n-1$ feels, beside the gravitational pull $T_{n-1}$, a torque from $n$ which slows it
down and a torque from $n-2$ which speeds it up. Generally the equation for domino $i$
has the form
\begin{equation} \label{d3}
{d \omega_i \over dt} = T_i +f_{i-1} a_{i-1} - f_i b_i.
\end{equation} 
The coefficients of the torques follow from the geometry shown in Fig. \ref{dominoes}.
\begin{equation} \label{d4}
a_i = h \cos (\theta_i - \theta_{i+1}) - (s+d) \sin \theta_{i+1},  \quad \quad \quad 
b_i = h \cos (\theta_i - \theta_{i+1}).
\end{equation}
Note that the first equation (\ref{d2}) is just a special case with $f_n = 0$. Another
interesting features is that $a_i<b_i$. So $i$ gains less from $i-1$ than $i-1$ looses
to $i$. Therefore dominoes, falling concertedly, gain less angular momentum than
if they would fall independently. This will have a consequence on the 
application of conservation of angular momentum in the collision process. We come back 
on this issue in the next section.

We can eliminate the forces from the equation by multiplying (\ref{d2}) with $r_0=1$ and
the general equation with $r_{n-i}$ and chosing the values of $r_j$ such that
\begin{equation} \label{d5}
r_j  = r_{j-1} \, {a_{n-j} \over b_{n-j}}, \quad \quad (r_0 = 1), \quad \quad \quad {\rm or} 
\quad \quad \quad r_{n-i}  = \prod^{n-1}_{j=i} {a_j \over b_j}. 
\end{equation} 
Then adding all the equations gives
\begin{equation} \label{d6}
\sum_i r_{n-i} \left[{d \omega_i \over dt} - T_i \right] =  
\sum_i [\,f_{i-1} \,r_{n-i} \, a_{i-1} - f_i \, r_{n-i-1} \, a_i\, ] = 0.
\end{equation} 
Now observe that the recursion for the $r_j$ is identical to that of the $w_j$ as given in 
(\ref{a10}). With $r_0 = 1$ we may identify $r_j = w_j$. It means that if
we  multiply (\ref{d6}) with $\omega_n$ and replace $r_{n-i} \omega_n$ by $\omega_i$,
we recover the conservation of energy in the form
\begin{equation} \label{d7}
{d \over dt} {1\over 2} \sum_i \omega^2_i = \sum_i \omega_i \, T_i. 
\end{equation} 
It is not difficult to write the sum of the torques as the derivative with respect to time of the
potential energy, thereby casting the conservation of energy in the standard form. So if
conservation of energy holds, the elimination of the forces is superfluous.
However, equation (\ref{d6}) is more general and we use it  in the treatment of friction.

\section{The Collision, second version} \label{coltrain}

We have assumed that in the collision of the head of chain $n$ with the next domino $n+1$ 
conserves angular momentum. Having a more detailed picture of forces between the 
sliding dominoes we reconsider this assumption. In this section without friction 
and in Section \ref{friction} with friction.
The idea is that in the collision domino $n$, exerts a impulse on $n+1$ and vice versa with
opposite sign. In other words: one has to integrate the equations of motion of the previous
section over such a short time that the positions do not change, but that the velocities 
accumulate a finite difference. 
However, not only the jump in velocity propagates downwards, also the impulses have
to propagate downwards in order to realize these jumps.
Denoting the impulses by capital $F$'s, domino $i$ receives $F_i$ from $i+1$ and 
$F_{i-1}$ from  $i-1$. So we get for the jumps in the rotational velocity
\begin{equation} \label{e1}
\left\{ \begin{array}{rcl}
\omega_{n+1} (0) & = & F_n \,a_n, \\*[2mm]
w_1 (0)\, \omega_{n+1}(0) - w_0 (\theta_c)\,\omega_n (\theta_c) & = 
& F_{n-1} \, a_{n-1} - F_n \, b_n,\\*[2mm]
\cdots & =  & \cdots \\*[2mm]
w_{n+1-i} (0)\, \omega_{n+1} (0) - w_{n-i} (\theta_c)\,\omega_n (\theta_c) 
& = & F_{i-1} \, a_{i-1} - F_i \, \,b_i.
\end{array} \right.
\end{equation} 
The functions $a_i$ and $b_i$ are the same as those defined in (\ref{d4}).
If we would have $a_i = b_i$ we could add all equations and indeed find that the angular 
total angular momentum is conserved in the collision. But only $a_n = b_n$ since 
$\theta_{n+1} =0$. The impulse $F_i$ can be eliminated in the same way as before by 
multiplying the $i$th equation with $r_{n+1-i}$ and adding them up. For the coefficient of 
$\omega_{n+1} (0)$ we get
\begin{equation} \label{e2}
\sum^{n+1}_i r_{n+1-i} \,w_{n+1-i} (0) = \sum^{n+1}_{j=0} r_j \, w_j (0) = J_{n+1}, 
\end{equation} 
and for the coefficient of $\omega_n (\theta_c)$ one finds with (\ref{a10})
\begin{equation} \label{e3}
\sum^n_i r_{n+1-i} \,w_{n-i} (\theta) = \sum^n_i r_{n+1-i} \,w_{n+1-i} (0) =
\sum^n_{j=1} r_j \, w_j (0) = J_{n+1} - 1.
\end{equation}
As general relation we get 
\begin{equation} \label{e4}
J_{n+1} \, \omega_{n+1} (0)  = (J_{n+1} - 1)  \, \omega_n (\theta_c).
\end{equation}  
In our frictionless case $r_j = w_j$ and therefore $J_{n+1} = I_{n+1}(0)$. So 
the desired relation reads
\begin{equation} \label{e5}
I_{n+1} (0) \, \omega_{n+1} (0)  = (I_{n+1} (0) - 1)  \, \omega_n (\theta_c) = 
I_n (\theta_c) \, \omega_n (\theta_c). 
\end{equation} 
We have added the last equality since it smells as a conservation of
angular momentum using the effective angular moment of inertia $I (\theta)$. This inertia
moment is however linked to the energy and not to the angular momentum. The true angular
momentum conservation is given in Section \ref{collision}. It is also not conservation of
kinetic energy. Then the squares of the angular velocities would have to enter.
The difference with the earlier relation
(\ref{c3}) is that the sum involves the squares of the $w$'s. This has a notable influence
on the asymptotic velocity. 

\section{Fully Developed Solitons} \label{solitons}

After a sufficient number of rotations and collisions a stationary state sets in. Then we
may identify in the collision law the entry $\omega_{n+1} (0)$ with $\omega_n (0)$. This
allows to solve for the stationary $\omega_n (0)$. We use (\ref{a11}) to relate
the effective moments of inertia
\begin{equation} \label{z1}
I_n(\theta_c) = \sum^n_{j=0} w^2_j (\theta_c) = \sum^{n-1}_{j=0} w^2_{j+1} (0) + 
w^2_n (\theta_c) = I_n (0) - w^2_0 (0) + w^2_n (\theta_c).
\end{equation}
For large $n$ the last term vanishes and we may drop the $n$ dependence in $I_n$. So
\begin{equation} \label{z2}
I (\theta_c) = I(0) - 1.
\end{equation} 
The collision laws thus may be asymptotically written as, 
\begin{equation} \label{z3}
I(0) \, \omega_n (0) = [\, I(0) -1\, ] \, \omega_n (\theta_c).
\end{equation}
The rotation is governed by the conservation of energy, which we write as
\begin{equation} \label{z4}
I(\theta)\, \omega^2_n (\theta) + H_n (\theta) = I (0)\, \omega^2_n (0) + H_n(0).
\end{equation} 
We can use (\ref{a9}) to relate the height function $H_n (\theta_c)$ to its value at 
$\theta =0$.
\begin{equation} \label{z5}
H_n (\theta_c) = \sum^n_j [\cos p_j (\theta_c) + {d \over h} \sin p_j (\theta_c)] =
H_n (0) -1 + \cos p_n (\theta_c) +  {d \over h} \sin p_n (\theta_c).
\end{equation} 
The limiting value of $p_n$ is the stacking angle $\theta_\infty$ 
Therefore the difference between the initial and the final potential energy reads
\begin{equation} \label{z7}
H (0) - H (\theta_c) = 1 - \cos \theta_\infty - {d \over h} \sin \theta_\infty \equiv P(h,d,s).
\end{equation} 
We have introduced the function $P$ as the loss in potential energy in the soliton motion.
It is the difference between an upright domino and a stacked domino at angle 
$\theta_\infty$. The functional form reads explicitly
\begin{equation} \label{z8}
P(h,d,s) = {s h  - d (s^2 + 2 s d) ^{1/2} \over h (s+d) }.
\end{equation} 
It is clear that the domino effect does not exist if $P$ is negative, because a domino tilted
at the stacking angle has a higher potential energy than an upright domino.

We use (\ref{z7}) in the conservation law for the energy, taken at $\theta = \theta_c$
\begin{equation} \label{z11}
I (\theta_c) \, \omega^2_n (\theta_c) - I(0) \, \omega^2_n (0) = P(h,d,s).
\end{equation} 
Solving $\omega_n (0)$ and $\omega_n (\theta_c)$ from (\ref{z3}) and (\ref{z11}) yields
\begin{equation} \label{z12}
\omega^2_n (0) = P(h,d,s) \, {I(0) - 1 \over I(0)} , \quad \quad  \quad  
\omega^2_n (\theta_c) =  P(h,d,s) \, {I(0) \over I(0)-1}.
\end{equation}
By and large $\sqrt{P}$ sets the scale for the rotation velocity. The dependence on $I(0)$
is rather weak. For large $I(0)$ it drops out. The minimum value of $I(0)$ is 2 which is
reached for large separations. 

\section{Friction} \label{friction}

After all this groundwork it is relatively simple to introduce friction. Let us start with
the equation of motion (\ref{d3}). Friction adds a force parallel to the surface
of $i+1$. For the strength of the friction force we assume the law of Amonton-Coulomb
\cite{amonton}
\begin{equation} \label{f1}
f_{\rm friction} = \mu f,
\end{equation} 
where $f$ is the corresponding perpendicular force. Inclusion of friction means that the
coefficients $a_i$ and $b_i$ pick up a frictional component. The associated torques follow
from the geometry of Fig. \ref{dominoes}. So the values of the $a_i$ and $b_i$ change to
\begin{equation} \label{f3}
\left\{ \begin{array}{rcl}
a_i & = & h \cos (\theta_i - \theta_{i+1}) -  (s+d) \sin \theta_{i+1}-\mu d, \\*[2mm]
b_i & = & h \cos (\theta_i - \theta_{i+1}) + \mu \, h \sin (\theta_i - \theta_{i+1}).
\end{array} \right.
\end{equation}
Then we may eliminate the forces as before, which again leads to (\ref{d6}). But we cannot 
identify any longer $r_i$ with $w_i$. In order to use (\ref{d6}) we must
express the accelerations $d \omega_i / dt$ in the head of chain $d \omega_n / dt$.
This follows from differentiating (\ref{a8})
\begin{equation} \label{f5}
{d \omega_i \over dt} = w_{n-i} (\theta_n) {d \omega_n \over dt} + v_{n-i} (\theta) \, 
\omega^2_n,
\end{equation} 
with $v_i$ given by
\begin{equation} \label{f6}
v_j (\theta) = {d w_j (\theta) \over d \theta_n}.
\end{equation}
The $v_j$ can be calculated from the recursion relation, that follows from differentiating
(\ref{a10}). Clearly the recursion starts with $v_0=0$ (see (\ref{f5})). 

Next we insert (\ref{f5}) into (\ref{d3}) and obtain
\begin{equation} \label{f7} 
\left(\sum^n_j r_j \,w_j \right)\, {d \omega_n \over dt} = \left(\sum^n_j r_j \,T_{n-j}\right)
-\left( \sum^n_j r_j \, v_j \,\right) \omega^2_n.
\end{equation}
The equation can be transformed into a differential equation for 
$d \omega_n / d \theta_n$ by dividing (\ref{f8}) by $\omega_n = d \theta_n / dt$ 
\begin{equation} \label{f8}
\left(\sum^n_j r_j \,w_j \right)\, {d \omega_n \over d \theta_n} = 
\left( \sum^n_j r_j \, T_{n-j} \right) {1 \over \omega_n} -\left( \sum^n_j r_j \, v_j \,\right) 
\omega_n.
\end{equation} 
We use this equation to find $\omega_n$ as function of $\theta_n$ and then (\ref{b6})
again to calculate the duration of the time between two collisions.

The inclusion of friction in the collision law is even simpler, since relation (\ref{e4}) remains
valid, but now with the definitions (\ref{f3}) for $a_i$ and $b_i$.

\section{Thin Dominoes}

\begin{figure}[h]
\begin{center}
    \epsfxsize=12cm
    \epsffile{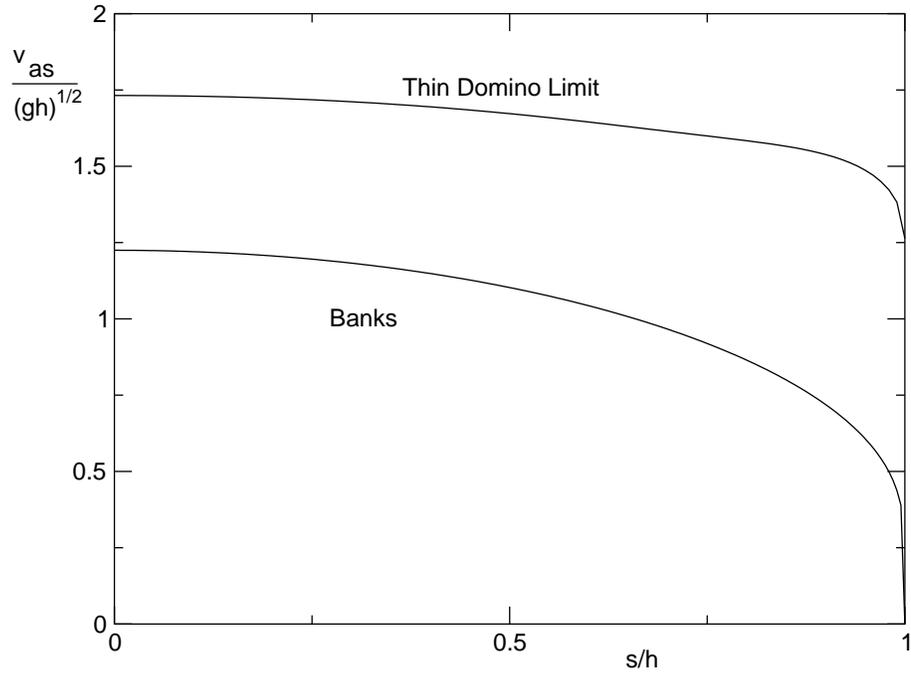} 
    \caption{The asymptotic soliton velocity as function of the separation $s/h$ in the thin 
domino limit. Also is plotted the result of Banks in the same limit.}  \label{cthans}
\end{center}
\end{figure}

Sometimes limits help to understand the general behaviour. One of the parameters,
which has played sofar a modest role, is the aspect ratio $d/h$. In our formulae it is
perfectly possible to take this ratio 0. In practice infinitely thin dominoes are a bit weird,
because with paperthin dominoes one has e.g. to worry about friction with the air.
In this limit we can vary $s/h$ over the full range from 0 to 1. 
In Fig.~\ref{cthans} we have plotted the asymptotic velocity as function of the separation $s/h$.
The curve is rather flat with a gradual drop--off towards the large separtions. We discuss 
here the two limits where the separation goes to 0 and where it approaches its maximum
$s=h$. Both offer some insight in the overall behavior.

\subsection{Infinitesimal Separation}

If the dominoes are narrowly separated, the head of chain rotates only over a small angle 
$\theta_c = \arcsin(s/h) \simeq s/h$ and the collisions will rapidly succeed each other.
The number of dominoes with a tilt angle $\theta_i$ between 0 and $\pi/2$ becomes 
very large and slowly varying with the index $i$.
So a continuum description is appropriate. We first focus on the dependence of 
$\theta_i (\theta_n)$ on the index $i$ and later comment on the dependence on the weak
variation with $\theta_n$ (which is confined to the small interval $0< \theta_n < \theta_c$).
We take as coordinate $x$ the distance of domino $i$.
\begin{equation} \label{h1} 
x = i \, s/h
\end{equation} 
and use $\nu = n s/h$ for the position of the head of the train. Then 
\begin{equation} \label{h2}
\theta_i = \theta(x), \quad \quad \quad  \theta_{i+1} = \theta (x +dx),
\end{equation}
with $ dx = s/h$. So for $d=0$ and $s/h \rightarrow 0$ the constraint (\ref{a1}) becomes
\begin{equation} \label{h3}
\sin [\,\theta(x) - \theta(x + dx)\,] = dx \cos \theta(x + dx),
\end{equation} 
leading to the differential equation
\begin{equation} \label{h4}
{d\theta(x) \over dx } = -\cos \theta (x),
\end{equation} 
which has the solution 
\begin{equation} \label{h5}
\sin \theta (x)  = \tanh (\nu - x) \quad \quad \quad {\rm or} \quad \quad \theta (x)  =
\arcsin (\tanh (\nu - x)).
\end{equation}
Here we have used the boundary condition that $\theta(\nu) =0$. Not surprisingly we
find that the shape of the tilt angles is a function of the difference with respect to the head
of the train. The above expression gives the shape of the soliton. 

Next we comment on the dependence of this profile on the angle $\theta_n$. As mentioned it
can be only weak as the interval for $\theta_n$ is narrow. Thus it suffices to know a few
derivatives and for that, the interpretation (\ref{a10}) is useful. The behavior
of $w_j$ in the continuum limit, follows from the differential form of the recursion relation
\begin{equation} \label{h6}
{d w(x) \over dx} = \sin \theta (x)\, w(x), 
\end{equation}
with the solution
\begin{equation} \label{h7} 
w(x) = {1 \over \cosh (\nu - x)} = \cos \theta(x). 
\end{equation}
Note that, not unexpectedly, the form of $w(x)$ follows also from that of $\theta (x)$ by 
differentiation with respect to $\nu$. Similarly the expression for $v_j$, as
given by (\ref{f6}), can be obtained from differentiation of (\ref{h7}) with respect to $\nu$
\begin{equation} \label{h9}
v(x) = - {\tanh (\nu - x) \over \cosh (\nu - x)} = - {d w(x) \over d x}.
\end{equation} 

What still is needed is the propagation velocity of the soliton, or in the present language: 
how fast $n$ or $\nu$ moves with time. As the foremost domino rotates over a small angle
$\theta_c \simeq s/h$, the head of train covers the distance $s/h$ with the 
rotation velocity $\omega_n$. So the propagation speed equals $\omega_n$. 
As before, $\omega_n$ has to be distilled from the laws of rotation and collision. 
Since this section is mainly for illustration, we restrict ourselves to the frictionless case. 

In the collision law (\ref{e5}) we encounter
$\omega_n (\theta_c)$ and $\omega_{n+1} (0)$. Both are linked to 
$\omega_n (0) = \omega(\nu)$ by
\begin{equation} \label{h10}
\omega_n (\theta_c) = \omega (\nu) + {\partial \omega_n \over \partial \theta_n} s/h, \quad \quad
\quad \omega_{n+1} (0) = \omega (\nu) + {\partial \omega \over \partial \nu } s/h.
\end{equation} 
For the derivative with respect to $\theta_n$, we
can take advantage of the form (\ref{f8}) which directly gives this derivative. We use that
$r_i = w_i$ in the frictionless case. The sums can be performed explicitly in the continuum
limit using (\ref{h7}) and (\ref{h9})
\begin{equation} \label{h11}
\left\{ \begin{array}{rclcl}
\displaystyle {s \over h} \sum^n_j w^2_j & = & \displaystyle \int^\nu_0 \,dx \,{1 \over 
\cosh^2 (\nu - x)}& = & \tanh \nu, \\*[6mm]
\displaystyle {s \over h} \sum^n_j w_j \, T_{n-j}  & = & \displaystyle \int^\nu_0 \,dx\, 
{\tanh (\nu - x)  \over 2 \cosh (\nu - x)} & = & \displaystyle 
{\cosh \nu - 1 \over 2 \cosh \nu}, \\*[6mm]
\displaystyle {s \over h} \sum^n_j v_j \, w_j  & = & \displaystyle - \int^\nu_0 \,dx \,
w (x) \, {d w \over d x} & = & \displaystyle - {\tanh^2 \nu \over 2}.
\end{array} \right.
\end{equation} 
Therefore the equation for $\partial \omega_n / \partial \theta_n$ becomes 
\begin{equation} \label{h12}
\tanh \nu \, {\partial \omega_n \over \partial \theta_n} = 
{\cosh \nu - 1 \over 2 \cosh \nu } \,{1 \over \omega_n } + {1 \over 2} \tanh^2 \nu \, \,\omega_n.
\end{equation}
With (\ref{h10}) the collision equation has the form
\begin{equation} \label{h13}
\sum^n_j w^2_j (\theta_c) [\,\omega (\nu) + {\partial \omega_n \over \partial \theta_n} s/h \,] =
[1 + \sum^n_j w^2_j (\theta_c)]\, [\omega (\nu) + {\partial \omega \over \partial \nu } s/h \,].
\end{equation} 
Using (\ref{h11}) we get, to first order in $s/h$,
\begin{equation} \label{h14}
\tanh \nu  {\partial \omega_n \over \partial \theta_n} = \omega(\nu) + \tanh \nu 
{\partial \omega \over \partial \nu }. 
\end{equation} 
Next we substitute (\ref{h12}) and we obtain the following differential equation for 
$\omega (\nu)$
\begin{equation} \label{h15}
\tanh \nu {d \omega (\nu) \over d \nu}  = \omega (\nu) ({1 \over 2} \tanh^2 \nu - 1)  +
 {\cosh \nu - 1 \over 2 \cosh \nu} \, \, {1 \over  \omega (\nu) }.
\end{equation} 
This awful looking differential equation has a simple solution
\begin{equation} \label{h16}
\omega^2 (\nu) = {\cosh \nu \over \sinh^2 \nu}\left(\cosh \nu -1 - \log \cosh \nu \right).
\end{equation}
We have chosen the integration constant such that $\omega (\nu)$ vanishes for $\nu = 0$.
It starts as 
\begin{equation} \label{h17}
\omega (\nu ) \simeq \nu / 2 \sqrt{2} + \cdots, \quad \quad \quad \nu \rightarrow 0,
\end{equation} 
and it saturates exponentially fast to the value $\omega(\infty) = 1$ 
(leading to $v_{as} = \sqrt{3gh}$). Thus we have obtained
in the continuum limit a full and explicit solution. It may serve as an illustration for the
general discrete case.

\subsection{Maximal Separation}

On the other side, near maximal separation $s \rightarrow h$, also a simplification occurs.
Here the number of dominoes involved in the train is restricted to a few. The head
of the train rotates over almost $\pi/2$ before it strikes the next domino. So one comes close
to the picture of Banks \cite{banks} in which the toppling of the dominoes is a succession of 
independent events. There is however a difference resulting from the constraint (\ref{a1}). 
Inmediate {\it after} the collision, the dominoes $n$ and $n+1$ rotate with equal velocity
$\omega_n (\theta_c) = \omega_{n+1} (0)$. This is a consequence of the fact that after the 
collision, one still has $\theta_{n+1} = 0$. Thus we  find for the energy after the collision
\begin{equation} \label{h18}
e_{n+1} = 1 + 2 \omega^2_{n+1} (0).
\end{equation}
All other dominoes have fallen down and domino $n+1$ is still upright (the 1 in (\ref{h18})).
Once $n+1$ starts rotating, the value of $\omega_n$ rapidly drops down to 0. 
Inspecting recursion (\ref{a10}), with $p_0 (\theta_{n+1}) = \theta_{n+1}$ and 
$p_1 (\theta_{n+1})= \theta_n$, one sees that the factor 
$\cos (\theta_n - \theta_{n+1}) \simeq \cos(\theta_c)$ is very close to 0. The ratio approaches
\begin{equation} \label{h19} 
{s \, \sin (\theta_{n+1}) \over h \, \cos (\theta_n - \theta_{n+1})} \rightarrow 
{s \, \sin (\theta_{n+1}) \over h \, \cos (\pi/2 - \theta_{n+1})} = {s \over h}.
\end{equation} 
So $w_1 \rightarrow 0$ and indeed domino $n$ comes to a halt; is has to, since it has reached the floor. This has an effect on the moment of inertia $I (\theta_{n+1})$ defined in
(\ref{b4}). Inmediately after the collision the value of $I (\theta_{n+1})$ equals 2, being the
sum of $w^2_{n+1} =1 $ and $w^2_n = 1$. A small angle further it has dropped to 1, 
since $w_n$ drops to 0.
As the energy is conserved the kinetic energy of domino $n$ is transferred to domino $n+1$.
So $\omega_{n+1}$ rises by a factor $\sqrt{2}$ in a short interval. Therefore we start the
integration of the time after this sudden increase, using the conservation law for the energy 
\begin{equation} \label{h20}
\omega^2_{n+1} (\theta_{n+1}) + \cos \theta_{n+1} = 1 + 2 \omega^2_{n+1} (0).
\end{equation}
In particular we have the relation for $\theta_c \simeq \pi/2$
\begin{equation} \label{h21}
\omega^2_{n+1} (\theta_c) = 1 + 2 \omega^2_{n+1} (0).
\end{equation}
The collision law for this degenerate case becomes
\begin{equation} \label{h22}
\omega_{n+1} (0) = \omega_n (\theta_c)/2.
\end{equation} 
Note that since the $w_j$ are either 1 or 0, there is no difference between the proposal
by Shaw (see (\ref{c3})) and ours (see (\ref{e5})). 

The stationary state is obtained by the identification 
$\omega_n (\theta_c)= \omega_{n+1} (\theta_c)$.
Combining (\ref{h21}) and (\ref{h22}) then yields $\omega_{n+1} (0) = 1/ \sqrt{2}$. Thus the time
integral for the interval becomes in the stationary state
\begin{equation} \label{h23}
t = \int^{\pi/2}_0 d \theta {1 \over \sqrt{2 - \cos \theta}} = 1.37 
\end{equation}
The reciprocal yields the asymptotic soliton speed $v_{\rm as} = 0.73*\sqrt{3} = 1.26$. 

The story of thin dominoes gives a warning on the numerical integration scheme. For small
separations we need many intervals before the asymptotic behaviour has set in. On the other
hand we do not need many points in the integration for the time of a rotation. For wide 
separations it is the opposite: only a few intervals are needed for the asymptotic behavior,
but we have to perform the time integration with care. The factor $I (\theta)$ in the 
energy law is rapidly varying for small $ \theta$. So we need many points for small 
$ \theta$ to be accurate. 

In Fig.~\ref{cthans} we have also plotted the curve due to Banks \cite{banks} in the limit of 
thin dominoes. The difference is due to the collision law, for which Banks takes 
\begin{equation} \label{h24}
\omega_{n+1} (0) = \cos \theta_c \, \omega_n (\theta_c).
\end{equation} 
The factor $\cos \theta_c$ accounts for the horizontal component of the linear momentum.
For large separations this gives quite a different value, since the transmission of linear
momentum becomes inefficient. For small separations the transmission is nearly perfect 
(as becomes our collision law).
The conservation of energy of a single rotating domino reads
\begin{equation} \label{h25}
\omega^2 (\theta) + \cos \theta = \omega^2 (0) + 1.
\end{equation} 
For the stationary state we insert (\ref{h24}) into (\ref{h25}) and get 
\begin{equation} \label{h26}
\omega^2 (\theta_c) = {1 - \cos \theta_c \over 1 - \cos^2 \theta_c}.
\end{equation}
For small $\theta_c$, the value $\omega(\theta_c)$   approaches $1/\sqrt{2}$, which is
again substantially smaller than our limiting value 1. The reason is that the dominoes,
which keep leaning onto each other and onto the head of the train, speed up the soliton.

\section{Calculations and Limitations} \label{calc}
\begin{table}[h] 
\begin{center}
\begin{tabular}{|r|c|c|c|c|c|} 
\hline
$s/h$ &Shaw & frictionless & $\mu=0.1$ & $ \mu =0.2$ & $\mu = 0.3$ \\*[1mm]
\hline
        &           &     &   &    &   \\*[-3mm]
0.1   &    3.64568   & 2.23469   & 1.82095  & 1.51695  & 1.28221 \\*[1mm]
0.2   &    3.06742   & 1.95534   & 1.66019  & 1.43423  & 1.25452 \\*[1mm]
0.3   &    2.74686   & 1.82515   & 1.56987  & 1.37279  & 1.21498 \\*[1mm]
0.4   &    2.50849   & 1.74231   & 1.50459  & 1.32193  & 1.17605 \\*[1mm]
0.5   &    2.30183   & 1.67865   & 1.44771  & 1.27272  & 1.13420 \\*[1mm]
0.6   &    2.10337   & 1.62447   & 1.39204  & 1.22009  & 1.08609 \\*[1mm]
0.7   &    1.89899   & 1.57824   & 1.33347  & 1.15958  & 1.02745 \\*[1mm]
0.8   &    1.68680   & 1.53779   & 1.26196  & 1.07970  & 0.94711 \\*[1mm]
0.9   &    1.47984   & 1.47984   & 1.15267  & 0.95822  & 0.82681 \\*[1mm]
\hline      
\end{tabular}
\end{center}
\caption{The asymptotic soliton speed ($d/h = 0.179$) for the collision law of Shaw and for
various degrees of friction with the collision law (\ref{e4}).} \label{tab}
\end{table} 
For the frictionless case we can use the formulae of Section \ref{solitons}, i.e. first
calculate the asymptotic value of $\omega (0)$ and then integrate the rotation equation
to find the time between two collisions and thus the asymptotic soliton speed. With friction
we must iteratively find $\omega (0)$, by trying
a value of $\omega (0)$, then solve equation (\ref{f8}) for 
$\omega  (\theta_c)$ and finally apply the collision law in order to see whether we come back
to our trial $\omega (0)$. A form of iteration is to start the train with
one domino and an arbitrary initial $\omega_0 (0)$ and let the train grow longer such
that an asymptotic pattern develops. 

\begin{figure}[h]
\begin{center}
    \epsfxsize=12cm
    \epsffile{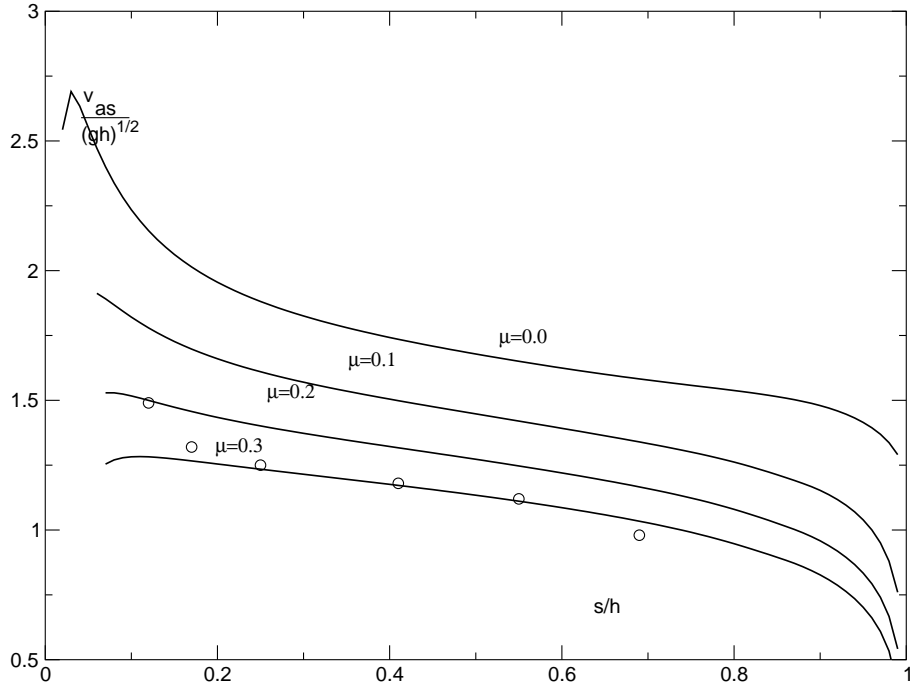} 
    \caption{The influence of friction on the asymptotic soliton speed for the aspect ratio
$d/h=0.179$. The dots are 
the experimental values of Maclachlan et al. \cite{maclachlan}.}  \label{survey}
\end{center}
\end{figure}

In Table \ref{tab} we have summarized the results. The thickness to height ratio is set
at $d/h = 0.179$ since this is the only value on which experiments \cite{maclachlan}
are reported. The first column gives the separation $s/h$, the second the soliton speed using 
Shaw's collision law 
and the third gives the results for ours (\ref{e5}). In the subsequent columns the 
influence of the friction is indicated. Note that the reduction of the speed due to the change
of the collision law is larger than that of modest friction. The
curves corresponding to these values are shown in Fig.~\ref{survey}, which also
contains the experiments of Maclachlan et al. They suggest that the soliton speed 
diverges for short distances, while we find a maximum. Their values seem to correspond 
best with the friction coefficient $\mu = 0.3$. We found empirically the value
$\mu = 0.2$,  by estimating the angle of the
supporting table at which dominoes start to slide over each other.

\begin{figure}[h]
\begin{center}
    \epsfxsize=12cm
    \epsffile{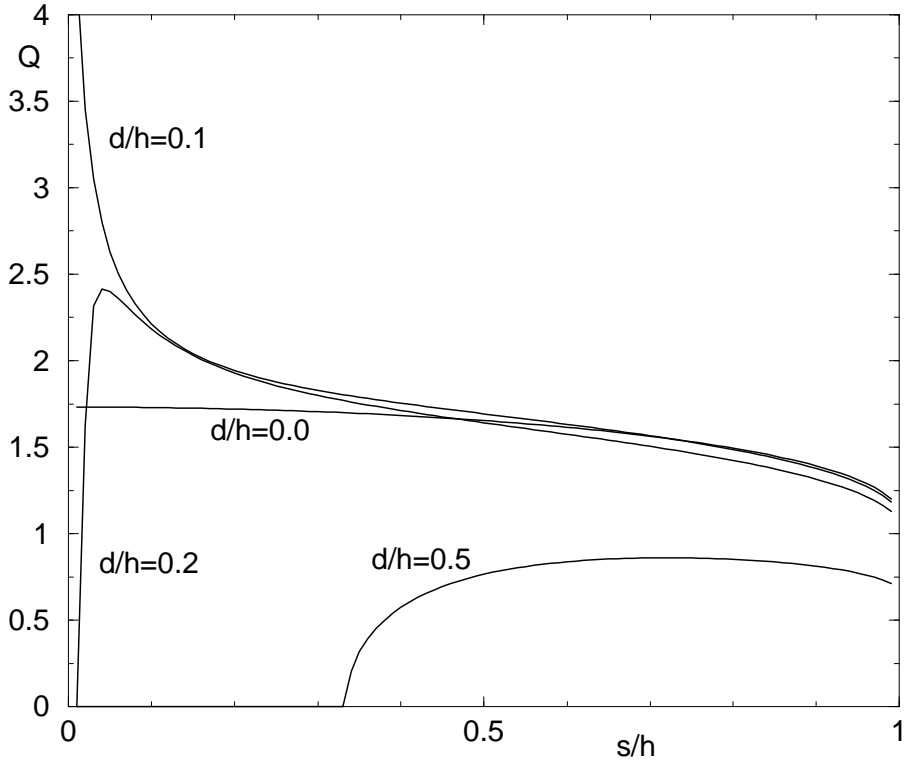} 
    \caption{The function $Q(s,d,h)$ as defined in (\ref{z15}) for various $d/h$.}  \label{shape}
\end{center}
\end{figure}

In order to make the behavior of fully developed solitons more transparant, we may 
introduce, for frictionless dominoes,  the average
\begin{equation} \label{z13}
{1 \over \langle \omega \rangle}  = {1 \over  \theta_c } \int^{\theta_c}_0 \, 
{d \theta \over \omega (\theta)},
\end{equation} 
with $\omega (\theta)$ the solution of (\ref{z4}).
This average is a number close to $1/\sqrt{P}$ (with $P$ defined in (\ref{z8})),
since the integrand varies from a value slightly larger than $1/\sqrt{P}$ to a value slightly
less than $1/\sqrt{P}$. Then we get for the asymptotic soliton speed the formula
\begin{equation} \label{z14}
{v_{\rm as} \over \sqrt{gh} } =Q(h,d,s) \, {\langle \omega \rangle \over \sqrt{P(h,d,s)}},
\end{equation} 
where the factor $Q$  is given by
\begin{equation} \label{z15}
Q(h,d,s) =\left({3\over 1 + d^2/h^2} \right)^{1/2} \,{(s+d)\,\sqrt{P(h,d,s)}\over h \arcsin(s/h)}.
\end{equation} 
Here we have reinstalled the factor $I/mgh$ in order to include in this formula, all the
factors that contribute to the velocity.
The factor $Q$ is shown as function of $s/h$ for various $d/h$ in Fig. \ref{shape}.
One may consider $Q$ as the main factor determining the dependence of the soliton speed on the
parameters of the problem. The fraction in (\ref{z14}) is a refinement which requires a 
detailed calculation. We found that this fraction is virtually independent of the aspect ratio 
$d/h$. It stays close to 1 for the the major part of the range of practical separations. 
Only around the already ``unworkable'' separation $s/h = 0.9$ the value has increased some 10\%. 
A good indicator for the behavior is the curve for the frictionless thin dominoes which is the 
product of the fraction and $Q=\sqrt{3} \,s/(h \arcsin(s/h)$. 

We mentioned in Section \ref{solitons} that the function $P$ as given by (\ref{z8}) has to
be positive for the existence of the domino effect.
This gives a bound on the minimal distance $s/h$, which can be cast in the form
\begin{equation} \label{z9}
{s \over h} > {2 (d/h)^3 \over 1 - (d/h)^2}.
\end{equation} 
Separations smaller than the value of (\ref{z9}) do not show the domino effect and slightly 
above that limit the train has difficulty to develop. The reason is that after a while,
too many dominoes of the train get tilt angles, which have a higher potential energy
than an upright domino. Ultimately the fraction of these dominoes in the train looses
out against the dominoes at the end of the train, which are tilted at the stacking angle
(with a potential energy {\em lower} than an upright domino). One can overcome this barrier
by starting with an unreasonble high initial $\omega_0 (0)$. So (\ref{z9}) is the true
theoretical limit, but in practice the domino effect will not start for slightly larger values 
of $s/h$.

Another limitation of the {\em theory} is at the other side. The dominoes at the end of the train
are tilted at the stacking angle $\theta_\infty$ {\em provided} the height $h$ is 
sufficiently large. The condition is 
\begin{equation} \label{i1}
h^2 > (s+d)^2 - d^2.
\end{equation}
For smaller $h$ the dominoes fall flat on the supporting table. (\ref{i1}) is satisfied for
\begin{equation} \label{i2}
s/h < \sqrt{1 + (d/h)^2} - d/h.
\end{equation}
Beyond this value the train is actually shorter than blind application of the formulae would
suggest. It is not so interesting to sort out what precisely happens if (\ref{i2}) is violated,
since then the no-slip condition for the dominoes is highly questionable. For such
wide separations the force on the struck domino has hardly a torque to rotate it. It rather 
induces the rotation axis to slide along the table. In fact, as a practical limitation, we look
to the height of impact. If it is above the center of mass of the struck domino, it will start to
rotate and below that value, it may slip if the friction with the supporting table is not
large enough. This criterion yields the limit to the distance
\begin{equation} \label{i3} 
s/h < \sqrt{3} \, / 2 = 0.87,
\end{equation}  
which is already a large separation, not far from the limit set by (\ref{i2}) for $d/h= 0.179$.

For very thin dominoes (\ref{z9}) is hardly a limitation. However,
(\ref{z9}) and (\ref{i3}) form a window of separations for the existence  of the domino effect,
which depends on the thickness $d/h$.
This window narrows down to zero and the domino effect disappears for
\begin{equation} \label{i4}
h^3 < h d^2 + 4 d^3 / \sqrt{3}~,  \quad \quad \quad {\rm or} \quad \quad \quad d/h < 0.3787
\end{equation} 
This estimate comes close to the one given by Freericks \cite{freericks}.
Friction also makes the excluded interval larger. For $\mu =0.2$ we have not found a domino
effect for $s/h < 0.07$, which is, for $d/h = 0.179$, about 7 times the theoretical
limit. So our estimate (\ref{i4}) for the upper thickness is still too optimistic. 

\section{Discussion}

We have studied the toppling of a row of equally spaced dominoes under the 
assumptions that the dominoes only rotate and that they keep leaning onto each other after a 
collision with the next one. By and large we follow the treatment of Shaw \cite{shaw}, 
who introduced the constraint (\ref{a1}), which  synchronizes the motion of train of
toppling dominoes. By analyzing the mutual forces between the
dominoes, we have corrected his collision law and  we could also account for the effect of 
friction between the dominoes. The correction of the collision law is more 
important than the influence of friction, given the small friction coefficient between
dominoes. The limit of thin dominoes $d/h \rightarrow 0$ leads to a completely tractable 
model. For large separations we encounter a situation which 
resembles the viewpoint of Banks \cite{banks}, seeing the toppling as a succession
of independent events. However his collision law differs substantially from ours and
cannot be reconciled with the force picture that we develop. We give a formula
(\ref{z14}), which displays explicitly the main dependence of the soliton speed on the 
parameters of the problem. The maximum speed which can be reached, appears close
to the closest separation for which the domino effect exists.

The assumptions, on which our calculation are based, are the no-slip condition and the
constraint (\ref{a1}). One can help the no-slip condition by increasing the friction with 
the supporting table (putting them on sandpaper as Walker \cite{walker} does).
If the no-slip condition is violated, it is the end of the domino effect as the dominoes are
kicked over with the wrong rotation. We argue that this will happen when the falling
domino hits the next one below its center of mass.

The constraint (\ref{a1}) is implied by the assumption
that the collision is fully inelastic. This assumption is supported by slow motion pictures
of the effect, which show that the dominoes indeed lean onto each other while falling.
It is an interesting question what happens, if the collision would be less inelastic.
The extreme opposite, fully elastic collisions, yields an ever increasing soliton speed.
A falling domino increases its rotation velocity as soon as its center of mass goes down. 
If this is fully transmitted to the next domino, the rotation velocity keeps increasing.
In that case friction can not play a role since the dominoes do not touch each other.
In the less extreme case of partially inelastic collisions, the dominoes also rotate without
contact, but friction can play a role during the collision. As Fig. \ref{dominoes} indicates,
friction always rotates the mutual impulse such, that the torque on the next one decreases and
that the reaction torque increases. This will slow down the train and a stationary state
can develop. Therefore it would be interesting to experiment with
dominoes of different making (e.g. steel) to see the increase in the soliton speed.

{\bf Acknowledgement}. This study was motivated by a question from the Dutch National 
Science Quiz of 2003 (www.nwo.nl/quiz). The author is indebted to Carlo Beenakker for 
drawing his attention
to the problem, for supplying the relevant literature and for stimulating discussions.


\begin{thebibliography}{99}
\bibitem{walker} Jearl Walker, Scientific American, August 1984.
\bibitem{banks}  Robert B.~Banks, {\em Towing Icebergs, Falling Dominoes and other adventures in 
Applied Mechanics},
Princeton University Press, 1998.
\bibitem{shaw}  D.~E.~Shaw, Am. J. Phys. {\bf 46} (1978) 640.
\bibitem{amonton} See e.g. D.~Tabor, ASME Journal of Lubrication Technology {\bf 103}
(1981) 169.
\bibitem{freericks}  J.~K.~Freericks, 
http://www.physics.georgetown.edu/~jkf/class\_mech/demo1.ps.
\bibitem{maclachlan} B.~G.~MacLachlan, G.~Beaupre, A.~B.~Cox and L.~Gore,
Falling Dominoes, SIAM Review {\bf 25} (1983) 403.
\end{thebibliography}
\end{document}